# REX: Recursive, Delta-Based Data-Centric Computation[*]

Svilen R. Mihaylov        Zachary G. Ives        Sudipto Guha

University of Pennsylvania
Philadelphia, PA, USA

{svilen, zives, sudipto}@cis.upenn.edu

## ABSTRACT

In today's Web and social network environments, query workloads include ad hoc and OLAP queries, as well as *iterative algorithms* that analyze data relationships (e.g., link analysis, clustering, learning). Modern DBMSs support ad hoc and OLAP queries, but most are not robust enough to scale to large clusters. Conversely, "cloud" platforms like MapReduce execute chains of batch tasks across clusters in a fault tolerant way, but have too much overhead to support ad hoc queries.

Moreover, both classes of platform incur significant overhead in executing iterative data analysis algorithms. Most such iterative algorithms repeatedly *refine* portions of their answers, until some convergence criterion is reached. However, general cloud platforms typically must reprocess *all* data in each step. DBMSs that support recursive SQL are more efficient in that they propagate only the changes in each step — but they still *accumulate* each iteration's state, even if it is no longer useful. User-defined functions are also typically harder to write for DBMSs than for cloud platforms.

We seek to unify the strengths of both styles of platforms, with a focus on supporting iterative computations in which *changes*, in the form of *deltas*, are propagated from iteration to iteration, and *state* is efficiently updated in an extensible way. We present a programming model oriented around deltas, describe how we execute and optimize such programs in our REX runtime system, and validate that our platform also handles failures gracefully. We experimentally validate our techniques, and show speedups over the competing methods ranging from 2.5 to nearly 100 times.

## 1. INTRODUCTION

In recent years, we have seen a major change in large-scale data analysis, in terms both of the kinds of data processing platforms being used and analyses being performed. Clusters of commodity hardware are used to extract structure and information from ever larger and richer datasets (Web or purchase logs, social networks, scientific data, medical records) — using increasingly sophisticated queries for relationship analysis, entity resolution, clustering, and recommendation. This has spurred the development of a new generation of scalable "NoSQL" cluster data processing platforms that analyze data outside the DBMS. Examples include Google's MapReduce [8] and its open-source Hadoop alternative, Pregel [20], Dryad [16] and Pig [21]. Such platforms are often targeted at large-scale "cloud computing" tasks, hence we refer to them as cloud platforms, even though they can be targeted at traditional clusters.

These cloud platforms share many implementation techniques with parallel DBMSs, yet they make different trade-offs that yield several benefits: scale-up to many (possibly heterogeneous) nodes, easier integration with user-defined code to support specialized algorithms, and transparent handling of failures (which frequently occur in large clusters). On the other hand, most platforms sacrifice one or more of: high-level programming abstractions, predefined primitives like joins, or declarative optimization techniques. Moreover, these platforms tend to be most useful for batch jobs, and some are extremely specialized (e.g., to graph processing [18, 20]). We make two key observations about such systems.

**Data analysis tasks increasingly need database operations as well as iteration.** Sources or relations may first need to be joined together, before being processed using algorithms for link analysis, recommendations, clustering, or learning graphical models. This requires a combination of database-style operations like joins, and support for iterative (or equivalently, recursive) processing. In each recursive step, many of these algorithms *refine* subsets of the answers until some convergence criterion is met. As we describe further in the next section, such computations are only inefficiently captured by recursive SQL queries, which *accumulate* but do not update answers. Most general-purpose cloud platforms do not directly support recursion or iteration — and must be repeatedly invoked, which is inefficient and does not take advantage of the fact that only *some* of the state changes from iteration to iteration. Cloud platforms specialized to iterative graph processing (e.g., Pregel) do support recursion, but they still do not fully capture the notion of incremental refinement, and they lack many data analysis primitives needed for general data, such as joins.

**The same data is often queried many ways.** As described in [25], a significant portion of the query workload for a company such as Facebook is OLAP queries — with joins, aggregation, and composition — to support planning (such queries may be executed as batch jobs) and reports on advertiser campaigns (which might be ad hoc queries). Much of the **same data**, along with social network data, is fed into complex, iterative ranking algorithms that determine the best placement of ads. Ideally, all of the data would be stored in the same platform, but made accessible to jobs ranging from small, quickly executed ad hoc queries (well-supported by a DBMS), through complex iterative batch jobs (in a cloud platform).

---

[*]Funded in part by NSF grants IIS-0477972, IIS-0713267, CNS-0721541 and a gift from Google. We thank Carlos Guestrin for his feedback on the work.

Naturally, there is a significant interest in blending techniques from both the database and "cloud" communities: this can be done by placing everything on a Hadoop-style implementation (e.g., Hive [23]), a hybrid architecture with a single programming interface (HadoopDB [1]), or multiple SQL engines (Greenplum). Some focus has been on developing better primitives (e.g., HaLoop [4] for recursion) or higher-level programming layers (e.g., Pig). However, to the best of our knowledge none of these efforts fully address the issues above, in supporting ad hoc and batch queries with support for joins and efficient iteration. We argue that efficient iteration requires that nodes maintain the state that has not changed from one iteration to the next, and that they only compute and propagate *deltas* describing changes. Of course, in the presence of failures, the platform must also efficiently *recover* that state.

**Our focus: supporting iterative algorithms that converge.** Our goal is to improve the performance of large-scale data analysis, for computations such as the following.

EXAMPLE 1. *Consider a directed graph stored as an edge relation, partitioned across multiple machines by* vertexId. *We want to compute the PageRank value for each vertex in the graph. A vertex's PageRank is iteratively defined: it is the sum of the weighted PageRank values of its incoming neighbors. Intuitively, a given vertex "sends" equal portions of its PageRank to each of its outgoing neighbors. Each aggregates "incoming" PageRank to update its new PageRank score for the next iteration. The process repeats until convergence: e.g., no page changes its PageRank value by more than 1% in the last iteration.*

For typical web or social graphs, 20-30 iterations are needed for convergence. The number of vertices which update their PageRank gradually decreases in the later iterations. For instance, in the last 15 iterations only an average of 10% of the vertices might update their PageRank values in each iteration. With general-purpose cloud platforms like MapReduce, which rely on functional (hence stateless) abstractions, each iteration's computation must be applied not only to the *changed* results from the previous step, but rather *all* answers. Some runtime systems for MapReduce, like HaLoop, mitigate some of I/O costs by caching data across iterations — but the actual computation is still performed repeatedly.

In contrast, a SQL DBMS could compute PageRank in an incremental way using recursive queries. Each iteration would accumulate an updated value for the PageRank of each node, based on the PageRank values from the previous iteration. We can make this even more efficient by only propagating, in each iteration, those PageRank values that have changed by more than some small amount. The SQL approach seems beneficial versus the MapReduce model[1]. However, the DBMS still suffers from the problem that recursive SQL *accumulates* state and does not allow it to be incrementally updated and replaced. For PageRank, we only need the *last* PageRank score for each tuple, but a recursive query does not allow us to discard the prior scores when we update them. (One could use materialized views and updates rather than recursive queries, but then the efficiencies of pipelining will be lost.) These observations are not restricted to PageRank, and in fact apply broadly to a wide variety of algorithms that perform optimization, random walks, and so on. Consider another example from clustering.

EXAMPLE 2. *Assume we have a set of two-dimensional points with real-valued coordinates and wish to cluster using* k-means clustering: *the result should be a small set of centroids, chosen such that the majority of the original points are as close as possible to (or clustered around) a centroid. To achieve this, we sample the initial centroid coordinates randomly among the coordinates of the given points, then assign each point to its closest centroid. We then re-assign the coordinates of the centroids as the average of their closest points, and repeat the re-assignment and re-computation steps until in the end no points switch centroids.*

Depending on the distribution of the point coordinates, it may take as many as 40 to 50 iterations for k-means clustering to converge. The number of points that switch centroids is large initially, but then gradually decreases as the centroid distribution begins to approximate the underlying point distribution. Similarly to PageRank, we would like to exploit this convergence behavior and iteratively propagate only the changes between iterations.

**The REX system.** In response to these requirements, we developed the REX system, which includes: (1) support for high-level programming using declarative SQL, (2) the ability to do pipelined, ad hoc queries as in DBMSs, (3) the failover capabilities and easy integration of user-defined code from cloud platforms, and (4) efficient support for *incremental* iterative computation with arbitrary termination conditions and *explicit creation of custom delta operations and handlers*. REX runs efficiently on clusters. Its generalized treatment of streams of incremental updates is, to our knowledge, unique, and as we show experimentally, extremely beneficial. REX supports extended SQL with user-defined Java code, and performs cost-based optimization. We take the further step of allowing the programmer to *directly use compiled code for Hadoop* in our environment: while adding some performance overhead versus a native approach, it provides an easy transition.

More specifically, our contributions are the following:

- A programming model and query language, RQL, with a generalized notion of *programmable deltas* (incremental updates) as first-class citizens, and support for user-defined code and arbitrary recursion.

- A distributed, resilient query processing platform, REX, that optimizes and executes RQL, supporting recursion with user-specified termination conditions.

- Novel delta-oriented implementation of known algorithms (PageRank, single-source shortest-path, K-means clustering) which minimize the amount of data being iterated over.

- Extensive experimental validation of the performance of the system on varied real-life workloads.

Section 2 reviews cloud programming platforms and their relationship to SQL DBMSs. Section 3 describes our contributions in the RQL language and model. Section 4 presents the high-level system architecture of REX. Section 5 describes how REX takes into account user-provided and inferred information when optimizing queries to produce an optimal execution plan. Section 6 experimentally validates the performance of REX by testing it under diverse conditions against a variety of other methods. Section 7 describes related work, and we conclude in Section 8.

## 2. QUERYING IN CLOUD PLATFORMS

This section briefly reviews the key concepts and constructs from cloud programming platforms and SQL, and shows how they are closely related. We first recall the basics of the programming model used by MapReduce [8], which is fairly representative of most of the other general-purpose platforms. MapReduce has achieved significant traction in the Web data analysis community due to its simple (if highly regularized) programming model, ability to scale to large commodity shared-nothing clusters, and failure resilience. MapReduce's programming and execution model are intertwined in a way that distinguishes them from database techniques.

---
[1]Note that the underlying query engine would also need to provide MapReduce-like levels of failure resilience.

**The MapReduce model.** MapReduce focuses on filter-aggregate computations and divides them into (up to) three stages: a user-specified *map* function that operates in parallel fashion across data that has been partitioned across nodes in a cluster, and that retrieves, filters, and specifies a grouping attribute for data items; an implicit *shuffle* stage that uses a sort-merge algorithm to group the output of the map stage; and a final user-specified *reduce* stage that performs an aggregation computation over the set of items corresponding to a single key. For performance reasons, an optional user-provided *combiner* may be invoked before the shuffle stage to pre-aggregate all results at a single node, before they are sorted. The basic MapReduce pipeline can be thought of as a special case of a single-block SQL GROUP BY query where the output is materialized. The mapper is equivalent to a user-defined WHERE condition and a specifier for which attributes to GROUP BY. The reducer is equivalent to a user-defined aggregate function. (The underlying execution model diverges from that of a DBMS: results are not pipelined across the network, but rather written to temp files that can be used as checkpoints for failure recovery.)

For simple computations, a MapReduce programmer need only specify a mapper and a reducer function, making basic programming very straightforward. Unfortunately, as data processing computations get more complex, they require coordination across multiple MapReduce jobs that combine and restructure data through a sequence of files. Even more complex are computations that require explicit iteration or recursion; such computations need to be managed by external control logic. Note that the MapReduce runtime has high startup cost, hence it is oriented towards batch jobs.

**Higher-level MapReduce abstractions.** Multi-step computations with composition and combination (i.e., join) become increasingly complex to write, maintain, and execute efficiently — and many real-world applications, e.g., those of Facebook as described in [25], require both joins and multiple aggregation steps. This has motivated the development of higher level, compositional languages that compile down to MapReduce or similar runtime systems, such as Pig Latin [21], DryadLINQ [26] and HiveQL [23]. Beyond composition, a major feature of these languages is join and nesting operations; however, none of these languages supports recursion with an arbitrary termination condition. As with basic MapReduce, performance is most appropriate for batch jobs.

**SQL-99.** SQL has many of the features desired in a high-level MapReduce abstraction: composition, joins, and even support for recursive queries. One missing feature is support for collection-valued attributes, which are essential to certain kinds of user-defined aggregations operations. Interfacing with user-defined functions is often cumbersome and proprietary to each DBMS, adding to the learning curve. Finally, SQL's support for recursion assumes that we are *accumulating* answers, whereas in many data-centric computation scenarios we are *refining* the answers.

As we describe in the next section, we seek to combine — then go significantly beyond — the best attributes of both cloud and DBMS platforms, with a focus on efficient support for refining answers, particulary for recursive queries.

## 3. RQL: SQL PLUS STATE MANAGEMENT

REX adopts a core declarative programming model, RQL, that is derived from SQL. However, it also seeks to minimize the learning curve for a non-database programmer who wishes to use the platform without knowing SQL. Hence REX can directly use Java class and jar files without requiring them to be registered using SQL DDL. It can directly execute arbitrary Hadoop MapReduce jobs consisting of multiple (possibly staged) map and reduce functions, for which it supplies a RQL query "template."

### 3.1 The RQL language

Of course, for more sophisticated programs, the software developer will need to write queries. REX's RQL language provides support for standard SQL features like joins, aggregation, and nested subqueries, as well as extensions for recursion, native Java objects (including collection-typed objects), and seamlessly embedded Java user-defined code.

By default REX supports a *stratified* model of recursion (though we are currently exploring other models). Execution begins with the base case, which forms the initial stratum. Once the base case has completed, the recursive case will be executed in its own stratum, combining the output of the base case with any other inputs. Each successive computation of the recursive case will only start after the previous stratum has completed. This allows REX to execute distributed computations in a way that produces deterministic results, as opposed to depending on the query author and algorithm to ensure meaningful results with asynchronous communication.

Traditional SQL is based on the notion of deriving relations from other relations (in the recursive case, the "other" relation may be an alternative version of the same relation). Over time, however, more general semantics emerged for taking a query defined purely over relation state, and automatically reasoning about how the output of the query should be updated, given changes to the query's input relations. For instance, *delta rules* [12] explain how tuple insertions or deletions get propagated from one relational operator to another, and also how they affect the internal state of "stateful" operators like joins and aggregation. Such concepts lie at the foundation of incremental view maintenance techniques, but also stream and continuous query [6] systems. As we shall discuss, we introduce a more general, **programmable** notion of delas.

### 3.2 State in REX

As in any DBMS, queries in REX's RQL language are optimized into query plans. REX supports the standard relational operators (**group by**, **join**, etc.) as well as a **while** operator (and its specialization, **fixpoint**) that governs the execution and termination of a recursive query, and an **applyFunction** operation that invokes a user-defined function over a tuple. (**Group by** and **join** also may call user-defined functions.) Finally, it is important to note that REX, like virtually all distributed query processors, has a physical level operator called **rehash** [15] that is responsible for shipping state from one node to another by key; this is essential, e.g., for supporting cross-machine joins.

In the relational algebra, several operators are *stateful* or *state-preserving*, in that they derive and buffer state based on the relation input they have received. We focus our discussion here on pipelined implementations of the operators, since those are generally preferable in a highly distributed context.

Consider a pipelined implementation of **group by** that maintains in its state a set of intermediate results for each grouping key. The actual form of the intermediate results will depend on the aggregate computation(s) being performed. This state will be accumulated until a *punctuation* marker — a special indicator of the end of a stream or a recursive step — is received, after which the operator will produce its output. Similarly, the **while** operator used for recursion will accumulate tuples produced from the current recursive step (stratum). After it receives a punctuation indicating the end of results for this stratum, it will propagate the new tuples to the next recursive step. Finally, the **join** operator, in its pipelined form, will accumulate each tuple it receives and immediately probe it against any tuples accumulated from the opposite relation — returning any joined tuples that can be produced from the probe.

Now let us consider how state is computed and accumulated in a standard recursive computation.

EXAMPLE 3. *Consider the PageRank example from earlier in the paper. We show the RQL query in Listing 1. We defer full details to later, but focus on the lower half of the listing, which looks like a recursive SQL query. As one possible way of executing the query using standard database operators, suppose the PageRank graph is represented as an edge relation partitioned across nodes by source page ID. An example query plan could be that of Figure 1. The "base case" of the query, in the upper right corner feeding into the* **fixpoint**, *reads the graph and assigns an initial PageRank score of 1 to every page (using* **applyFunction***). The resulting table is fed to a* **fixpoint** *operator (a specialization of* **while***), which iterates until no new results are produced. This operator maintains* state, *namely a relation mapping from page ID to PageRank, which is fed back into the next recursive iteration through the fixpoint receiver (marked as* PR*) and subsequently mapped to the join bucket corresponding to the value of srcId (see the rectangle labeled* prBucket*). Once each node executing the base case completes execution, it sends a* punctuation *marker to the* **fixpoint** *operator, which is essentially a "vote" to advance to the next iteration or* stratum.

*Now any* changes *to that state made in the current iteration are propagated to the first invocation of the recursive part of the query. This recursive portion of the query joins with the original graph (whose state is visualized as* nbrBucket*). These results are accumulated and redistributed (rehashed) according to the IDs of the pages' link* destinations: *i.e., the PageRank scores are distributed to the target pages. These new PageRank scores get summed up and then are accumulated as new tuples in the* **fixpoint** *operator. As an optimization, one could add an* iteration count *attribute to distinguish among the different scores.*

Observe that while the query executes correctly in this setting, it wastes resources. The computation only needs the latest PageRank scores, but the various query operators accumulate *all* state. Thus, we want to update **while** operator's state, in particular revising the old scores with new values. Similarly, as updated scores get propagated from the **while** operator back to the next iteration of the recursive step, we want to update — not merely add to — the values stored in the state of the **join** and **group by** operators. The ability to replace, revise, or delete results that are being accumulated is what we refer to as *refinement* of state.

## 3.3 Refining State

In the relational query processing literature, the notion of update propagation has been formalized through the notion of *delta tuples* [12] describing insertions, deletions, or replacements of tuples. A delta consists of an *operation* (insertion, deletion, or replacement) and up to two tuples. The insertion delta specifies a new tuple value to insert; the deletion delta specifies an existing tuple value to delete; the replacement delta specifies an existing tuple to replace, plus a new tuple value. Standard rules exist (e.g., in [12, 17]) for how to propagate insertions, deletions, and replacements through state-preserving operators like **join** and **while**.

The **group by** operator has many subtleties: in general, an update will be applied to the group whose key is associated with the delta. However, the rules for what to propagate for that grouping key depend on the specific aggregate operations being applied to the group. For instance, a **min** aggregate will take a tuple deletion delta, and first determine whether the deletion affects the existing minimum value. If so, it must determine the next-smallest value (which needs to be in its buffered state), then output a replacement delta, replacing the "old" minimum value with the new one. Otherwise it merely updates the set of values it has buffered. Compare this to a **sum** aggregate, which will take a tuple deletion delta, subtract a value from the current sum, then propagate a replacement delta to update from old to new sum. Finally, consider **average**, which is often divided into two portions: a *pre-aggregate* operation that associates both a sum and a count with each group (called **combiner** in MapReduce), and a *final aggregate* operation that accumulates pre-aggregate results and finally computes an average.

We summarize the take-aways of the above discussion: (1) A **group by** operator's internal state includes a map from the grouping key to some aggregate function-specific form of intermediate state, for each aggregate function being computed. (2) As a **group by** operator receives a delta, it can determine the key associated with the delta, but then each aggregate function needs to determine how to update its own intermediate state and what to emit.

If we generalize from these examples to the broader class of algorithms used in cloud data management, then the set of potential aggregate functions is unbounded, since such functions are user-defined. Moreover, it is often more efficient to model not merely *insertions, deletions, and replacements* of tuples, but rather *adjustments* to aggregate values. To do this, we generalize the notion of deltas in RQL, to allow for **extensibility**.

REX's mechanism for specifying and propagating updates is that of tuples with *annotations*. Annotations specify how the tuples are to be processed with respect to state: as insertions, deletions, replacements, or — our new functionality — value-updates.

DEFINITION 1 (DELTA). *A delta in REX is a pair* $(\alpha, t)$ *where* $t$ *is a tuple and* $\alpha$ *is an* annotation *describing an operation.* $\alpha$ *can be one of* $\{+(), -(), \rightarrow (t'), \delta(E)\}$ *where:*

- $+()$ *represents that* $t$ *is to be* inserted *into an operator state*
- $-()$: $t$ *is to be* deleted *from an operator state (if it exists)*
- $\rightarrow (t')$: $t$ *is a replacement for tuple* $t'$
- $\delta(E)$ *specifies an arbitrary* expression code $E$, *which encodes operations and parameters that must be interpreted by downstream "stateful" operators. As we describe below* $E$ *will ultimately be interpreted by user-defined code, which in turn will apply an appropriate operation to any existing state corresponding to the item* $t$, *plus* $t$ *itself, to form an updated version of the state.*

**Deltas and stateless query operators.** Deltas are automatically propagated by stateless operators in a simple way. If one views the delta as a tuple with an annotation, then the operator processes the tuple in the normal fashion (possibly filtering or projecting the tuple). Any output tuples receive the same annotation as the input tuple. (One exception to this general rule is the **applyFunction** operator, which is stateless but can create or manipulate annotations in arbitrary ways.)

Stateful operators, however, must update their state in accordance with the deltas, and propagate the effects to the next stage.
**Join:** For insertions, deletions, and replacements, propagation follows the rules established in [12]. Conceptually each delta is considered in sequence. Insertions and deletions are applied, then probed against the state from the opposite relation, and the results are propagated as insertions or deletions. Replacements are effectively treated as deletion-insertion sequences, whose results may be deletions, insertions, or replacements. User-defined annotations can be combined using a *delta handler* for updating the join state, plus a second delta handler defining how to combine annotations from tuples that join, as described below.
**While:** Deltas are directly propagated, and the operator's internal state is modified in accordance with insertion, deletion, or replacement logic. User-defined annotations can be combined with existing state using a delta handler.

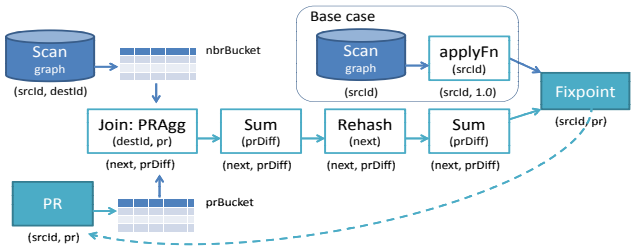

Figure 1: Query plan for computing PageRank

**Group by:** Each user-defined aggregate function has special rules on how state should be updated and what deltas should be propagated. The standard operators (min, max, sum, average, count) automatically handle insertion, deletion, and replacement deltas. For all other situations, the programmer must supply delta handlers.

As described above, for a variety of situations, the user must specify code defining how to process a tuple with a given annotation. If such a delta handler is not provided, REX will propagate the annotation as if it were another (hidden) attribute of the tuple, with no special semantics. There are four forms of delta handlers:

- **aggregate state**: DELTA[] AGGSTATE(OBJECT STATE, DELTA D). Called by the **group by** operator with the existing state associated with the grouping key (a default object if the key does not exist), and revises the state object according to the new delta tuple. It may optionally return a delta representing an intermediate value, e.g., a partial sum.
- **aggregate result**: DELTA[] AGGRESULT(OBJECT STATE). Called by **group by** with the existing state after the stratum has finished to return a final sequence of deltas.
- **join state**: DELTA[] UPDATE(TUPLESET LEFTBUCKET, TUPLESET RIGHTBUCKET, DELTA D). Called by a **join** operator with the corresponding joining tuple buckets. It can modify the buckets according to the input delta, and generate resulting delta tuples.
- **while state**: DELTA[] UPDATE(TUPLESET WHILERELATION, DELTA D). Called by a **while** operator and returns a new set of tuples, possibly the empty set.

From the above, it should be clear that "aggregate functions" in RQL are more than simple SQL functions: rather, they have two or more handlers defining how they manage and propagate state. We refer to such entities as *user-defined aggregators* or UDAs, and they are a distinguishing feature of our programming model. UDA methods (and other kinds of user-defined functions) are strongly typed and typechecking is performed by the query processor. Moreover, the base datatypes map cleanly to Java types, since the engine internally uses Java objects and scalar types to represent its data. Even with the strong typing, user-provided classes in REX have broad flexibility in their implementation. Programmers may directly instantiate a typed Java interface (e.g., a user-defined type providing an aggregate method over a sequence of integers, returning an integer); they may simply provide a set of methods with designated names reserved by the system whose typing information is retrievable by Java reflection; or they may provide metadata, also queried by reflection, about input and output parameter types.

### 3.4 Deltas and Recursion

Many data-centric programming models (SQL, MapReduce variants) require an *explicit termination*: e.g., the PageRank value for two consecutive iterations of every node needs to be checked for convergence. Accordingly, REX allows the user to join or otherwise compare the recursive output from different strata to compute explicit termination conditions: *How many pages have their Page-Rank changed by more than 1% between iterations n and n-1?*

**Listing 1: Computing PageRank with REX**

```
class PRAgg {
  String[] inTypes = {"Integer", "Double"};
  String[] outTypes = {"nbr:Integer", "prdiff:Double"};
  Object[][] update(Object[][] prBucket,
    Object[][] nbrBucket, int nbrId, double pr) {
    double deltaPr = prBucket.get(nbrId)-pr;
    prBucket.put(nbrId, pr);
    if (Math.abs(deltaPr) > 0.01) {Object[][] resBag={};
      for (Integer nbr: nbrBucket)
        resBag.add(nbr, deltaPr/nbrBucket.size());
      return resBag;
} } }
WITH PR ( srcId, pr) AS        -- Base case initializes ...
( SELECT srcId, 1.0 AS pr FROM graph -- PageRank to 1
) UNION UNTIL FIXPOINT BY srcId ( -- Recursive case ...
  SELECT nbr, 0.15+0.85*sum(prDiff) -- produces *deltas*
  FROM ( SELECT PRAgg(srcId, pr).{nbr, prDiff}
        FROM graph, PR  -- deltas from prev. iteration
        WHERE graph.srcId = PR.srcId GROUP BY srcId)
  GROUP BY nbr)
```

An alternative type of recursion in REX is *fixpoint recursion* with *implicit termination*, which was heavily studied for Datalog. Here, execution of a query terminates once the current stratum is complete, with no observable change to its output. Often explicit recursion can be recoded as implicit recursion (see the leftmost column of Figure 3 for examples). This model, with its emphasis on propagation of deltas, generally results in much more efficient computation than approaches that re-process the data in each iteration as it changes (we term this the *mutable set* of a computation). We refer to the set of changed tuples for iteration $i$ of the query as the $\Delta_i$ set. The $\Delta_i$ set represents the minimal set of tuples required to be processed for a given iteration. The $\Delta_i$ set abstraction is applicable to a range of distributed algorithms in the domains of machine learning and graph analysis. Figure 3 summarizes the types of data for several other distributed algorithms. The appendix has implementations for some of those.

### 3.5 PageRank in RQL

Let us now revisit the PageRank example of Listing 1. The top half is Java code that declares a user-defined join handler named *PRAgg* for the relations *PR* and *graph*, and a recursive RQL query which uses it. For readability, the Java syntax has been simplified. inTypes and outTypes aid REX in determining the types of the input and output parameters, by providing SQL attribute names.

Looking at the query, the *base case* simply assigns a default weight to each query. The recursive step consists of a nested subquery dividing PageRank among all out-edges, followed by an aggregation that sums up all "incoming" PageRanks. The nested subquery applies the *PRAgg.update* function (specified in the SELECT clause) to each (join) pairing of graph and PR tuples satisfying *graph.srcId = PR.srcId*. Looking at the Java code, this function updates the state objects from the *PR* and *graph* relations with matching join keys, here called *prBucket* and *nbrBucket*. PRAgg is also called with two further arguments representing the user-defined delta arguments (deltas are encoded in curly braces in the query): here, these are the page ID and a diff to apply to the PageRank. (The operation is implicit, namely an arithmetic sum.) The Java logic states that, if the change in PageRank value is sufficiently large (greater than 1%), multiple delta tuples are generated in output result *resBag*. Each neighbor receives an equal fraction of the change of the current PageRank value. The query continues until we reach a (set-semantics) fixpoint, where we use the *srcId* to divide the accumulated results into partitions.

The set of pages annotated with their current PageRank value in each recursive step form the *mutable set* discussed in the previous section. Each changes in every iteration. In contrast, the graph

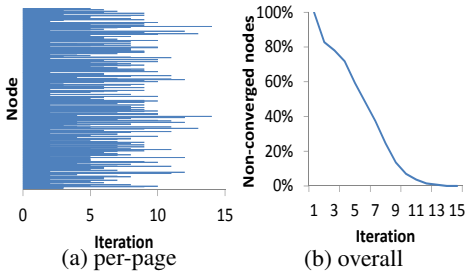

| Algorithm | Immutable set | Mutable set | $\Delta_i$ set |
|---|---|---|---|
| PageRank | graph edges | PageRank value for all vertices | PageRank values with change $\geq$ 1% since iteration i-1 |
| Adsorbtion | graph edges | complete adsorbtion vectors for all vertices | adsorbtion vector positions with change $\geq$ 1% since iteration i-1 |
| Shortest path | graph edges | minimum distance for reachable vertices | vertices with minimum distance from source at iteration i lower than their distance at iteration i-1 |
| K-means clustering | coordinates | full assignment of nodes to centroids | nodes which switched centroids at iteration i |
| CRF learning | document set | model parameters | parameters updated at iteration i |

(a) per-page    (b) overall

Figure 2: PageRank convergence behavior

Figure 3: Types of recursive data

itself does not change during the computation and thus we refer to it as the *immutable set*.

As the computation proceeds, the PageRank values will begin to converge and thus the differentials will shrink. As the differentials shrink, fewer and fewer nodes will issue diffs to their neighbors until eventually no diff tuples are issued, leading to convergence. Figure 2 provides a sample of the convergence behavior of PageRank using $\Delta_i$ sets. Although individual pages require different number of iterations to converge to a stable PageRank value, the overall number of non-converged nodes steadily decreases.

## 4. STORAGE AND RUNTIME SYSTEM

REX is a parallel shared-nothing query processing platform implemented in Java, combining aspects of relational DBMSs and cloud computing engines. The input query is submitted to a *requestor* node which is responsible for invoking the RQL query optimizer and distributing the optimizer query plan and referred Java user-defined code (UDC) to the participating query "worker nodes". UDC runs in the same instance of the Java virtual machine as the Java code comprising the REX implementation and is invoked via the Java reflection mechanism. The query optimizer performs top-down optimization [10] as discussed in Section 5. After the query has been optimized by the requestor node, it is disseminated to all workers for execution.

The input data resides on partitioned replicated local storage. The system incorporates a data-partitioned parallel stream engine based on [22], but significantly extended to execute recursive queries with UDCs and support incremental recovery. The query processor seeks to maximize its use of main memory, but has the ability to spill overflow state to local disks as necessary, or to repeatedly scan or probe against disk-based storage. Each worker node executes in parallel the query plan specified by the optimizer. The results of the plan execution are ultimately forwarded to the query requestor node, which unions the received results from all nodes in the cluster. There is no single node responsible for checkpointing the state, coordinating flows, etc. of the query execution.

### 4.1 Communication and Replication

Architecturally REX differs significantly from MapReduce, and builds upon the basic scheme of [22]. Data partitioning is based on keys rather than pages, and partitions are chosen using a consistent hashing and data replication scheme known to all nodes. Communication is achieved via TCP with destinations chosen by partitions: there is no abstraction of a distributed filesystem, and query processing passes batched messages.

REX's data partitioning scheme ensures that every node can access any non-local data (possibly via a replica) without consulting a central directory server. Importantly, every query in REX is distributed along with a snapshot of the data partitions across the machines *as seen by the query requestor*. All data will be routed according to this set of partitions, guaranteeing that even as the network changes, data will be delivered to the same place. If a node fails, the *recovery* process will use the partition snapshot to determine what data needs to be recomputed by replica nodes. (During each recovery process, the data partition snapshot gets updated to reflect the new set of nodes.) See [22] for more details on how this is implemented and how failures are detected. When a coordinator is needed for validating correctness or ensuring distributed consensus, this task is taken on by the *query requestor* node: i.e., the node making a query request is responsible for coordinating it.

### 4.2 Distributed Pipelined Execution

As with many distributed query engines, execution in REX is data-driven, starting at the table scan operators reading local data and pushing it through the other operators (which are virtually all pipelined, including a pipelined hash join). All operators have been extended to propagate and handle deltas in the fashion described in Section 3. Selection and aggregation operators in REX are extended to handle user-defined code, and also cache results for deterministic functions. We also implement a variant of the *dependent join* that passes an input to a table-valued function and combines the results: this operator even supports calls to multiple table-valued functions in the same operation. Whenever needed, a rehash operator re-partitions data among worker nodes based on the partitioning snapshot for the current query.

The system supports input batching for UDC, which may take sequences of input tuples instead of one tuple at a time. This amortizes the overhead of Java reflection calls (needed to determine parameter types) across multiple function invocations, at the cost of a moderate increase in memory. Streamed *partial aggregation* operators are supported: partially aggregated table-valued output may occur after processing one or several input tuples. This can help to avoid maintaining large internal state, and is particularly useful when executing native Hadoop code as described in Section 4.4.

REX extends the traditional set of query operators with **while** and **fixpoint** operators. The fixpoint operator has a dual function: it forwards its input data back to the input of one operator in the recursive query plan, and also removes duplicate tuples according to a query-specified key, by maintaining a set of processed tuples.

Recall that REX supports implicit (no new tuples produced in the current stratum) and explicit (some arbitrary condition holds between two consecutive strata) termination conditions. REX can convert explicit termination condition into implicit one by executing the explicit condition into a separate subquery producing a boolean result, which in turn filters tuples from the main computation.

The REX engine uses punctuation [24] (special marker tuples) to inform query operators that the current stratum is finished. At the end of a stratum, all fixpoint operators send the **number** of processed tuples to the query requestor, which informs the operators whether the query implicit termination condition has been met (i.e., no new tuples have been computed in the current stratum). Depending on the result, either an *end-of-query* or *end-of-stratum* punctuation is sent to the recursive step of the query. Punctuation is propagated in a straightforward way: unary operators like selection or aggregation simply forward it directly to their parent operators, while n-ary operators such as a join or rehash wait until all inputs have received appropriate punctuation before proceeding.

## 4.3 Incremental Recovery for Recursion

In large-scale data-centric processing, machine failures (due to crashes, hardware failures, etc.) are so common that they are often a primary consideration. Arguably, one of the strengths helping MapReduce achieve widespread use is its simple and robust *checkpointing* mechanism: intermediate state or results are saved onto distributed file system and in the case of a failure the computation resumes using that state. On the other hand, common approaches for recovery in parallel DBs tend to either simply restart the query whenever a failure has been detected or use *recovery queries* [22] to re-compute portions of the final result affected by the failures. During recovery, the remaining nodes need to recompute all results from the failed node; moreover, they may need to replace any data that "passed through" operators on the failed nodes to prevent data duplication. This requires query operators to "tag" every tuple with the set of nodes involved in its processing, as described in [22].

Neither of those extremes is optimized for recursive computations: MapReduce (even when pipelined [7]) takes a pessimistic approach by essentially checkpointing *all* intermediate state, which can be expensive. Conversely, a traditional pipelined approach makes the optimistic assumption that no checkpointing is required — thus requiring a complete query restart or multiple recursive executions of recovery queries if a failure does occur. Of these two cases, the former guarantees progress under realistic assumptions, but incurs a performance penalty; whereas the latter has less overhead but does not guarantee forward progress.

REX employs a hybrid recovery approach (as suggested by [11]) combining the strengths of checkpointing and recovery queries. We employ *incremental checkpoints*: for a given stratum, every machine buffers and replicates the mutable $\Delta_i$ set processed by the local fixpoint operator to *replica machines*. In the presence of failures, recovery queries are started from the last stratum which was successfully completed, minimizing the time required for the recovery query to catch-up. The checkpointed tuples in the failed range are streamed to the nodes which have taken over that range, allowing the recursive computation to resume. During each successive recursive case, the number of newly derived tuples (the $\Delta_i$ set) typically stays constant or shrinks as we approach convergence, meaning the overall state in a typical setting remains manageable. Thus, the incremental recovery mechanism employed by REX is robust and efficient and, importantly, guarantees forward progress even in the presence of repeated failures.

## 4.4 Executing Hadoop Code in REX

To enable users to rapidly migrate their existing code, REX allows direct use of compiled code for Hadoop by utilizing specially designed table-valued "wrapper" functions. While this adds some performance overhead versus a native approach (incurred because of the requirement for Java reflection calls), it can still provide performance benefits. Hadoop jobs typically consist of a driver program which defines the inputs, outputs, and dataflow between map and reduce classes, and definitions of those map and reduce classes. A driver program for a single MapReduce job involving a map and a reduce class can be expressed with the following query:

```
SELECT ReduceWrap('ReduceClass',
       MapWrap('MapClass', k, v).{k, v}).{k, v}
FROM InputTable GROUP BY MapWrap('MapClass', k, v).k
```

*InputTable(k,v)* contains (key,value) tuples. *ReduceWrap* and *MapWrap* are specially written UDA and UDF which take key-value input tuples as well as an additional constant specifying a Reduce or Map class (named *ReduceClass* and *MapClass* on the example).

Chained or branched jobs can be expressed as nested subqueries within a compound driver query, each of which follows the same basic structure and corresponds to each job. Iterative execution of MapReduce jobs can similarly be achieved by constructing a query with a base case and a recursive case. Unlike examples presented earlier, the input and output types of *ReduceWrap* and *MapWrap* are not statically defined, but are instead automatically computed from the actual typed inputs of the keys and values used by the Hadoop classes. This helps REX eliminate the impedance mismatch between its internal types and those used in Hadoop code. Additional wrappers are provided to convert between native tuples to text or binary format suited for Hadoop code and vice versa.

As the results in the experimental section confirm, the REX platform is often able to execute native Hadoop code faster than the Hadoop framework. Though a detailed study is beyond the scope of this paper, the reasons for this advantage lie mainly in reducing the checkpointing requirements (section 4.3) by avoiding expensive disk IO after every job, as well as avoiding the relatively expensive disk-based external merge sort required by the shuffle phase.

## 5. OPTIMIZATION

In a traditional database system, the query optimizer relies on offline-computed data-distribution statistics (often histograms over important attributes) to estimate operator selectivities; from this it predicts plan costs and chooses the lowest-cost plan. Our setting is somewhat more complex, as there may be many worker nodes and the queries include aggregation, recursion, and user-defined functions. Leveraging ideas present in many distributed and parallel optimizers [10, 14, 19] REX's query optimizer uses top-down enumeration and branch-and-bound pruning. We consider a combination of network, disk, and CPU costs, and take into account different interleavings of joins and user-defined functions, the possibility of pre-aggregation, and the expected cost of recursive computation. In this section, we highlight some aspects of our query optimizer.

**Many-node cost estimation.** We assume that each node has run an initial *calibration* that provides the optimizer with information about its relative CPU and disk speeds, and all pairwise network bandwidths. From these values, the optimizer uses, for each operator, the *lowest* combined cost estimate across all nodes: this in essence estimates the worst-case completion time for each operation. Some operators like filescans are primarily disk-bound; others like rehash are primarily network-bound; and still others like user defined aggregator (UDA) calls are likely CPU-bound. REX uses both pipelining and multiple threads during query execution, and hence it is often possible to overlap I/O- and CPU-intensive tasks.

**Accounting for CPU-I/O overlap.** To help take potential overlap into account, REX models pipelined operations using a *vector of resource utilization levels*. Rather than simply adding the execution times to produce the overall runtime, the REX optimizer determines the result runtime as the lowest value that allows both subplans to execute in parallel while the combined utilization for any resource remains under 100%. In the extreme case where the two subplans use completely disjoint resources, the resulting runtime equals the maximum of the runtime of the subplans, rather than their sum.

**Query plans for deltas.** Deltas are represented as arbitrary parameterized annotations in the RQL programming model. During query optimization, they are translated into additional fields in the tuples, and this allows the optimizer to reuse its logic for determining constraints on operator ordering.

In the remainder of this section, we describe how REX's optimizer determines how to find the best interleaving of UDF calls and joins, determines whether to perform pre-aggregation, and optimizes recursive queries, respectively.

## 5.1 Ordering of UDFs and Joins

As has been noted in past work, in the presence of potentially expensive user-defined functions, the usual System-R heuristic of maximally pushing all selection or applyFunction predicates is often suboptimal. Instead, during plan enumeration, we must consider different interleavings of joins and UDFs to find a minimum-cost strategy. However, considering all orderings of UDFs leads to an exponential increase of the plan search space: In general there are $n!$ possible orders of application of $n$ independent predicates over the same relation, and even more possible ways to "merge" any such order with the rest of the query plan (including joins, rehashes, etc). We build upon the results of [13], which shows that the optimal order of application of expensive predicates over the same relation is in increasing order of *rank*, which is defined as the ratio of the cost per tuple versus the selectivity. Intuitively, this result states that predicates which are inexpensive to compute, or discard the most tuples, should be applied first. Our optimizer extends the approach of [13] by incorporating the UDFs in an execution model with overlapping CPU, disk, and network resources (as described above) and across bushy as well as linear join plans.

**Caching.** Functions can be marked as *volatile* or *deterministic*: for deterministic functions, REX will cache and reuse values, and it can take this into account during cost estimation.

**Cost calibration and hints.** REX uses a set of calibration queries plus runtime monitoring to estimate the per-input-tuple cost, running time, and selectivity or productivity of a UDF. Without knowing any semantics of the function, REX assumes that the cost is value-independent. However, certain classes of functions have costs dependent on their input values (which in turn may be known to the optimizer because they are literals in the query, or values known to be in the distribution). For such settings, we allow programmer-supplied *cost hints* — functions describing the "big-O" relationship between the main input parameters and the resulting costs (e.g., one value may control the number of iterations within the function). Such information provides REX with the basic "shape" of the cost function, which it combines with its calibration routines to determine the appropriate coefficient for estimating future costs.

## 5.2 UDA Pre-aggregation Pushdown

Many aggregate functions can be partly precomputed in advance, e.g., before sending data to another node (achieved via *combiners* in MapReduce), or before joining the results with another relation. If a UDA also supplies a *pre-aggregation* function, then REX's optimizer will also explore possible interleavings of pre-aggregation with UDAs and joins. We adopt the approach of [5], but impose the heuristic that only one pre-aggregation is allowed for each UDA, and that it is maximally pushed down. Unlike the prior work, for maximum flexibility we consider table-valued aggregations.

**UDA composability.** If UDAs are marked as *composable*, then they are computable in parts, which can be unioned together and a final aggregation can be applied (e.g., *sum* and *average* but not *median*). For such UDAs, we can push the pre-aggregation through any arbitrary join operation, whereas for non-composable predicates REX can only push them under a key-foreign key join. Importantly, when used in a query, any aggregate function is used as if it were composed into a single function with its pre-aggregate. REX performs composition taking into account typing information, thus hiding the pre-aggregation altogether from the query.

**Composability and multiplicative joins.** Some composable functions (e.g., sum) have a value dependent on the number of tuples in the aggregate set. There is a certain special case where we might wish to perform pre-aggregation on *both inputs* to a join that is not on a key-foreign key relationship. Here we would ordinarily have $m$ tuples for each group from the left input join with $n$ tuples from the group on the right — but if both are pre-aggregated, we will under-estimate the final result. If the user specifies an optional *multiply* function, REX will perform this pre-aggregation, and will compensate for the under-estimate by multiplying the inputs by the cardinality of the group on the opposite join input. For simple built-in numerical aggregates, the *multiply* function corresponds to actual numerical multiplication. For complex table-valued UDAs (possibly with non-numerical types) a *multiply* user-defined function (itself a UDF) must be specified by the user with the *multFn* field. The input type of the multiply function must be the same as the input of the aggregation function, with the addition of one integer argument corresponding to the cardinality of the other join group (count(*)). The output type should be the same as the input type of the aggregation function. The addition of the count(*) to the query plan is handled transparently by the optimizer.

## 5.3 Optimizing Recursive Queries

As we described in Section 2, recursive RQL queries consist of a *base case sub-query* and a *recursive sub-query* repeatedly unioned together using set semantics by a fixpoint operator. In effect, we simulate the repeated execution of the recursive query to optimize it. We start by considering the base case as a separate query block. Once we obtain the optimal plan for the base case as well as its estimated cardinality and cost, we then treat this result as the input to the recursive step, and optimize the *first iteration* of the recursive case using our estimates for the base case's output. To estimate the overall plan cost, we iteratively proceed as follows. We take the estimated output of the recursive case in the current iteration, treat this as an input into the next iteration, optimize the next iteration, and repeat. During each iteration, we always cap our estimate of the recursive input size to be *no larger* than the previous stage's input, as (1) our focus is on recursive algorithms that converge and (2) we expect that the Fixpoint operator will remove duplicate derivations. The process terminates once our estimated output size for an iteration reaches zero, or we reach a maximum number of iterations. Our final plan will be assembled by combining the base and recursive case plans under a fixpoint operator.

During the iterative approximation of the cardinality of the recursive query, the resulting cardinalities might diverge (e.g. each iteration multiplies the cardinality of the fixpoint relation by a factor of 2). To avoid such exponential growth (possibly caused by poor optimization hints), REX limits the cardinality and cost for a given approximation step to be no more than the respective cost and cardinality computed during the previous step.

## 6. EXPERIMENTAL VALIDATION

Our primary goal is to measure the relative performance of the competing platforms: REX, general-purpose cloud platforms, and where feasible, a commercial DBMS. Specifically, we aim to demonstrate the effectiveness of the Delta-based evaluation strategy and the overall performance and scalability of REX under varied conditions. We also demonstrate our system's resilience to failures and low-overhead execution of native Hadoop code.

**Algorithms.** To achieve data and computation diversity, we tested platform performance using the PageRank, shortest-path, and K-means clustering algorithms, as well as one non-recursive aggregation query. We measured the total time to compute a complete result set, and (where applicable) the time needed to complete iterations during the result computation. For select algorithms we also show the required bandwidth for their execution. For each data point in the following results we performed nine runs, excluding the first two and averaging the other seven. We include 95% confidence intervals on all plots (often too narrow to be visible).

**Data.** There are four datasets used in the experiments. For PageRank and shortest-path we used article linkage graph obtained from DBPedia (dbpedia.org) and a user-follower dataset obtained by crawling Twitter (law.dsi.unimi.it/datasets.php). After normalization and cleanup, the DBPedia dataset contains 48 million tuples which encode a directed graph with 48 million edges (article links) and 3.3 million vertices (articles). The Twitter dataset contains 1.4 billion edges (follower-user pair) and 41 million vertices (users). For k-means clustering we used a DBPedia dataset containing longitude and latitude for 328232 DBPedia articles, which we also enlarge by simulating up to 1000 additional points around each original coordinate (382 million tuples). Finally, in order to assess the overhead of UDFs versus built-in aggregation functions, we used the *lineitem* relation from the TPCH benchmark, generated for 10GB total size. The resulting table contains 60 million tuples.

**Platforms.** We compared REX queries against implementations in Hadoop-style MapReduce. Our goal was to do a fair, direct comparison against Hadoop, and also against the HaLoop extended version [4] that includes recursion support. However, due to a variety of technical issues HaLoop and Hadoop do not coexist on the same (in our case, shared multiuser) cluster infrastructure. Thus we installed the Hadoop runtime and configured our experiments in a way that emulates HaLoop. We took the technical description of HaLoop's techniques in [4] and for each technique they developed for improving efficiency, we counted it as executing in **zero time**:

- construction of the Reducer Input cache
- recursive MapReduce stages involving immutable data

Moreover, for algorithms as executed in *either* Hadoop and HaLoop, we assumed an idealized implementation, in which the following execute in zero time:

- convergence test after each iteration
- input and output data formatting, and final result collection

Thus our numbers for HaLoop and Hadoop represent **lower bounds** on their actual performance numbers under realistic scenarios.

While we did not have access to a cluster-based DBMS (e.g. Greenplum or Oracle RAC), we also performed a comparison of recursive computation in REX versus a commercial DBMS, which we refer to as DBMS X, on a single machine (Section 6.4). We also attempted to compare REX against Giraph, an open-source graph-based platform modeled after Pregel. Unfortunately, Giraph is currently in alpha release, and its large memory footprint and inability to spill data to disk, as well as configuration mismatches with our existing Hadoop install, prevented us from doing direct comparisons. We only successfully ran PageRank on Giraph with 16% of DBPedia dataset with lower performance than REX.

**Configurations.** We tested REX with several options: *delta*, *no-delta*, and *wrap*. Delta provides the finest control over (and minimizes) the iterated data. It iterates over portions of the mutable data which change between iterations. No-delta iterates over all of the available mutable data (like Hadoop and HaLoop), and *wrap* is a case where we execute native Hadoop Java code (mapper, reducer, and combiner classes) using special wrapper UDFs and UDAs responsible for formatting the input and output data as strings, as well as ensuring proper fixpoint semantics. In effect, *wrap* tests the performance of emulated Hadoop jobs within REX. Unlike Hadoop and HaLoop above, the timing of each of the three methods includes the overhead of the initial loading of the data and formatting the output. No-delta and wrap do not perform convergence testing, while delta is actually required to do so in order to achieve efficient execution. On performance graphs which show recursive queries Hadoop and HaLoop are annotated as *Hadoop LB* and *HaLoop LB* respectively to emphasize the fact that they represent a lower bound (best-case). *REX* methods are run with the ability to **incrementally recover** from failures should they occur. We also include specific experiments with actual node failures.

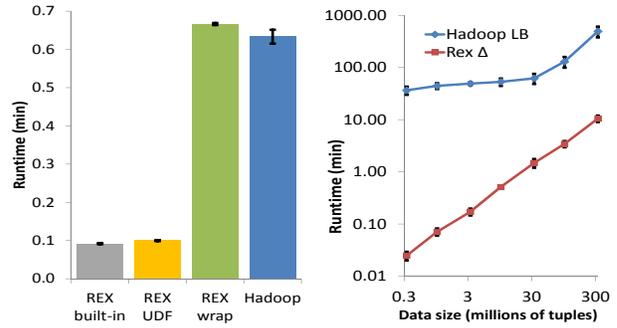

Figure 4: Simple aggregation   Figure 5: K-means clustering

**Cluster setup.** Experiments in this section were performed on a cluster of 28 quad-core 2.4Ghz Intel Xeon machines with 4GB of system memory running Fedora Linux. Java Hotspot/SE 1.6 was used as the default Java runtime environment, and Hadoop 0.20.1 was used for experiments involving Hadoop performance evaluations. The default settings of Hadoop were modified to allow the execution of 4 concurrent map or reduce tasks per machine in order to fully utilize the 4 available cores. 120 reduce tasks total (or 4.28 reduce tasks per machine) were specified for the execution of each Hadoop job, which is within the optimum range recommended by Hadoop developers: from 3.8 to 7 per machine in the instance of 4 available cores/machine. In other words, there are approximately as many reduce tasks allocated as the number of available cores.

### 6.1 UDF Overhead – Simple OLAP Query

Before testing recursive queries, we characterize the overhead of REX UDFs and UDAs versus built-in operators using a simple aggregation query over TPCH data:

`SELECT sum(tax),count(*) FROM lineitem WHERE linenumber>1`

Figure 4 shows the results of the execution of this query. *Built-in* specifies execution using only built-in operators: *sum*, *count*, as well as the *greater* selection predicate. For REX the two standard aggregations and selection are instead performed by 2 UDAs and 1 UDF. Hadoop shows execution of the same query in the Hadoop environment, using a single job consisting of one mapper, reducer and a combiner (necessary for efficient pre-aggregation). REX-wrap shows the execution of the Hadoop Java code using wrapper classes. Built-in and REX are faster than Hadoop by more than a factor of 3, showing that REX is much more appropriate for ad hoc or OLAP queries. Both REX and REX-wrap are no more than 10% slower than their native execution counterparts, thereby validating the efficiency of the UDC extensions in REX.

### 6.2 Deltas: Mutable-only Relations

Figure 5 shows the results of executing K-means clustering on the geographic DBPedia dataset for REX delta and Hadoop. We did not include HaLoop, since the query does not contain a relation with immutable data, meaning that HaLoop and Hadoop exhibit essentially the same behavior (as demonstrated by experiments in [4]). We tested the relative performance and scalability of both systems by varying the size of the input data from 382 thousand to 382 million tuples. REX delta is almost **two orders of magnitude** faster, due to its extremely low iteration overhead.

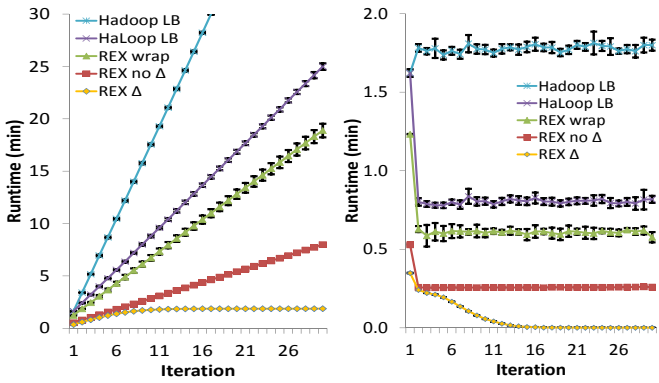
(a) Cumulative runtime  (b) Runtime per iteration
Figure 6: Recursive behavior for PageRank (DBPedia)

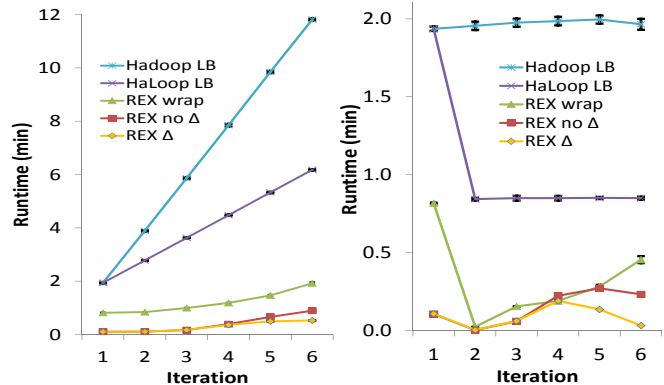
(a) Cumulative runtime  (b) Runtime per iteration
Figure 7: Recursive behavior for shortest-path (DBPedia)

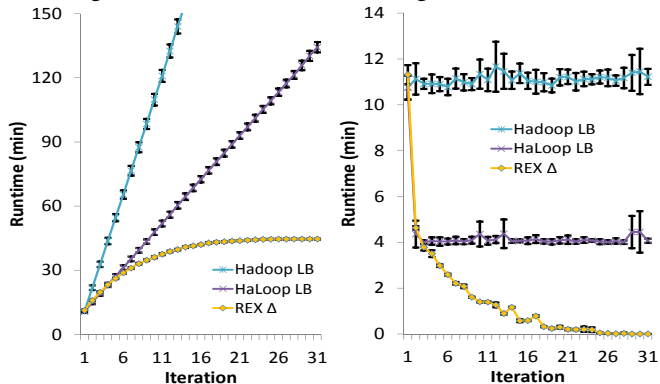
(a) Cumulative runtime  (b) Runtime per iteration
Figure 8: Recursive behavior for PageRank (Twitter)

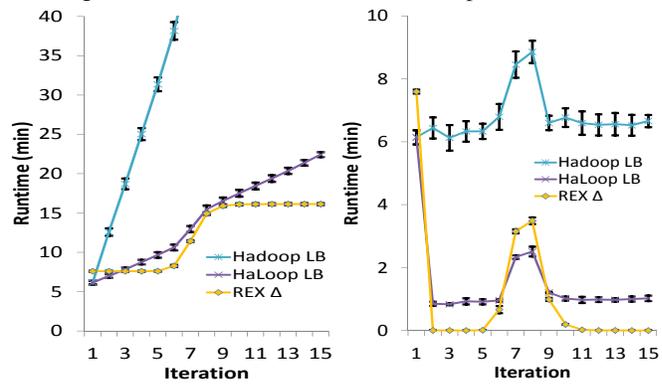
(a) Cumulative runtime  (b) Runtime per iteration
Figure 9: Recursive behavior for shortest-path (Twitter)

## 6.3 Deltas: Mutable and Immutable Relations

The benefits of REX are not limited to computations with mutable relations. In this subsection we focus on queries (PageRank and unweighed shortest-path) over the **DBPedia dataset**, showing two different aspects of our incremental computation model.

Figure 6(a) for the PageRank query reveals that REX delta is the best strategy for the DBPedia dataset, outperforming HaLoop by a factor of 10 and REX no-delta by a factor of 4. Figure 6(b) shows the runtime of all strategies excluding Hadoop and REX delta drop by nearly a factor of 2 after the initial iteration and stabilizing afterwards, as they always iterate over the entire set of nodes. However, the $\Delta_i$ set (over which REX delta iterates) steadily decreases, and the last iterations are performed progressively faster. But this set does not decrease monotonically since any given page might only intermittently meet the convergence criteria: e.g. less than 1% change of PageRank in iteration 2 but more than 1% change in iteration 3. Moreover, the $\Delta_i$ set is data dependent and no filtration strategy can discover it a-priori.

Figure 7 shows the execution of the shortest-path query on the DBPedia dataset. Unlike PageRank and K-means clustering, here it is possible to use a well-defined "frontier set" corresponding to a relation-level $\Delta_i$. (The frontier set initially consists of the starting node and expands outward by 1 hop for every iteration.) We have therefore ensured that both Hadoop and HaLoop use relation-level $\Delta_i$ updates for this query. Figure 7(a) shows REX delta being the dominating strategy, with a factor of 2 speedup over REX no-delta and nearly **an order of magnitude** over HaLoop.

**Improved Accuracy:** Although both graphs show execution of only six iterations, the diameter of the DBPedia graph is so large it requires 75 iterations to compute full reachability. For all methods except REX delta we perform only six iterations, enough to provide 99% reachability. REX delta itself performs all 75 required iterations, with iterations 7 to 75 taking under 1s in combined time. Note that all other methods require approximately as much time for each of the iterations from 7 to 75 as they do for iteration 6.

**Additional Benefits of REX:** Interestingly, REX-wrap is nearly twice as fast as HaLoop, even when both run the same computation. For recursive queries, the overhead of transforming the input data from text into format suitable for HaLoop and back into text is incurred only once in the beginning and in the end of the query. Thus we can expect that for recursive situations most of the overhead of REX-wrap (such as in Figure 4) can be avoided, and this is indeed true. Moreover, the architecture of REX avoids the expensive sorting step used in Hadoop and HaLoop and uses hash-based GROUP BY instead. This combines with the substantially lower start-up overhead of REX to produce significantly faster execution.

## 6.4 Scalability and the DBMS

We use the significantly larger **Twitter dataset** to compare the performances of the best possible alternatives: Hadoop, HaLoop, and REX delta. All the algorithms use relation level updates. Figure 8(a) shows that REX delta outperforms HaLoop by a factor of 3 and Hadoop by a factor of 7 when executing PageRank. Figure 8(b) shows the same phenomenon as Figure 6(b). Figure 9(a) shows that REX delta remains the faster strategy when it comes to computing shortest-path: it is faster than the lower bound HaLoop by nearly 30%. Figure 9(b) reveals a large jump in the per-iteration runtime around iterations 7 and 8, preceded and followed by very fast iterations. This is due an explosion in the size of the reachability set which occurs 7 hops from the initial node. The large spike in the first iteration reflects the time required to load the immutable data.

**Improvements vs. traditional DBMSs:** We show that REX also performs well when compared to an RDBMS. We configured DBMS

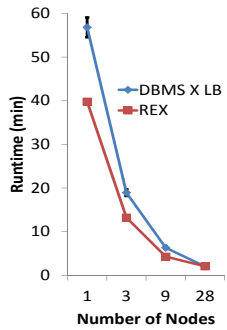 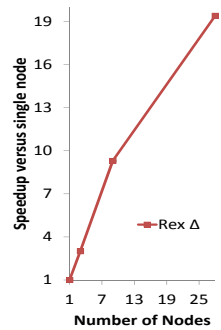 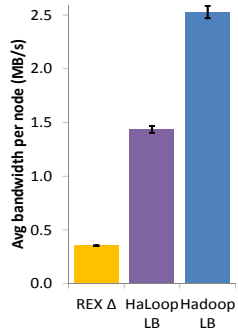 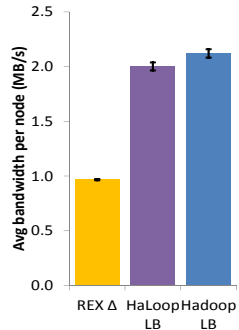 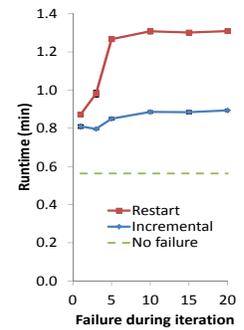

| (a) Scalability | (b) Speedup | (a) shortest-path | (b) PageRank | Figure 12: Recovery (DBPedia, shortest-path) |

Figure 10: Scalability and speedup

Figure 11: Avg. bandwidth usage (Twitter)

X in query-only mode with the smaller DBPedia dataset (to avoid disk configuration issues for the commercial DB engine). Figure 10(a) demonstrates the effect of varying the number of available machines on the time required to complete the PageRank query for the DBPedia dataset. As we increase the number of machines from 1 to 28, the runtime decreases proportionally. We also show PageRank on the same dataset using *DBMS X* on a single machine. We could not run the engine on multiple machines due to license issues, however the figure shows a lower bound on DBMS X assuming perfect linear speedup. On a single machine, REX delta performs about 30% faster than the commercial DB engine, and real-world REX performance always beats the idealized DBMS X. Figure 10(b) demonstrates the near-linear scalability of REX by showing the relative speedup versus execution on a single machine.

### 6.5 Bandwidth Utilization

Figure 11 shows the average bandwidth per node during the execution of each of the methods for the Twitter dataset. For *REX delta* we measured the total amount of data sent by each node and divided by the total number of nodes and the duration of the query. For Hadoop and HaLoop we aggregated the total amount of data shuffled per job, dividing by the number of nodes and duration of the query. *REX delta* achieves significantly lower bandwidth than either Hadoop or HaLoop in the case of PageRank: 0.97 MB/s versus 2.00 MB/s. For shortest-path, the difference is even more pronounced. This result shows that *REX delta* is the better choice in comparatively bandwidth limited environments such as P2P systems.

### 6.6 Recovery from Node Failure

In an ideal system, performance improvements should not compromise failure tolerance. Figure 12 shows execution time of the shortest-path query with the DBPedia data set in the presence of a node failure whose timing varies from iterations 1 to 20. The time to complete the query is recorded. Two competing options were tested: *Restart* represents the baseline with the query simply restarted when a failure is detected, discarding work completed prior to the failure. This strategy does not need to replicate the mutable data. *Incremental* on the other hand utilizes work done prior to the failure to reduce the time to complete the query. This strategy replicates the mutable data (in the case of shortest-path: the current minimum distances for each node) in the DHT. A replication factor of 3 is used. When a node failure occurs at iteration k, nodes which take over the failed range resume the execution without having to recompute the mutable data up to iteration k. As can be seen from the results, the incremental strategy halves the recovery overhead as compared with the case where no failures are present. Although we did not test under such extreme scenarios, the incremental strategy would allow forward progress even in the case of repeated failures, whereas the restart strategy will have to discard work and start the query from the beginning on every new failure.

### 6.7 Summary of Experimental Results

We conclude that *REX delta* is clearly the best strategy, consistently outperforming Hadoop and HaLoop independently of the size of the data and query setting. The Hadoop framework has a substantial startup and tear-down overhead, which particularly affects the K-means query, as well as those iterations of the shortest-path queries during which the reachability set is stable. *REX delta* often outperforms HaLoop by more than a factor of 3.

## 7. RELATED WORK

We have already discussed MapReduce, Hadoop and HaLoop in some detail, as well as Pregel, DryadLINQ, Pig, and Hive. Cloud platforms for iteration and graph processing have attracted significant attention. Spark [27] extends MapReduce with a distributed shared memory abstraction for preserving state, but does not address optimization. The paper [2] extends MapReduce to support incremental updates with a focus on minimizing effort to re-compute single-stage MapReduce jobs when their input changes. Twister [9] extends MapReduce to preserve state across iterations, but it only supports failure recovery through complete restarts. It also does not optimize non-recursive computation.

GraphLab [18] develops a programming model for *iterative* machine learning computations. Its model is specialized to learning, consisting of *update* functions where state for particular keys changes, followed by *sync* operations where (like reduce) results are globally aggregated. GraphLab includes primitives for sequential and parallel synchronization, but uses sequential scheduling and partitioned, lock-based shared memory.

Another branch of related work combines database and MapReduce models. MapReduce Online [7] introduces pipelining to MapReduce, while still spooling data to disk in parallel to enable recovery from failures. It supports continuous queries and incremental output, but does not address recursive queries. Hyracks [3] is a parallel dataflow engine with a pipelined execution model over a DAG of abstract operators; it can execute MapReduce-style jobs or query operators for XQuery. Given the acyclic nature of its plans, its focus is on non-recursive tasks. HadoopDB [1] is a hybrid of Hadoop (MapReduce) and PostgreSQL: queries are posed in a variant of SQL based on Hive, the basic core of Hadoop manages the computations, but the computations are partly "pushed" to PostgreSQL implementations running on each node. It supports aggregation and incremental recovery, but not recursion.

## 8. CONCLUSIONS

REX is an extensible, reliable, and efficient parallel DBMS engine that supports user-defined functions, custom *delta* updates, and iteration over shared-nothing clusters. It seamlessly embeds Java code within SQL queries, and provides flexible recursion with state management — thus supporting many graph and learning algorithms. The query processor includes optimizations specific to queries with user-defined functions, aggregates, and recursion. Our optimizer extensions do not come at the expense of performance, and special attention is paid to function reordering and pre-aggregation pushdown. The optimizer also supports user-provided hints to improve estimates. REX even supports execution and composition of native Hadoop code in the relational engine, thus subsuming the functionality of MapReduce.

### Listing 2: Computing shortest-path with REX

```
class SPAgg {
  String[] inTypes = {"Integer", "Double"};
  String[] outTypes = {"nbr:Integer", "distOut:Double"};
  Object[][] update(Object[][] distBucket,
    Object[][] nbrBucket, int nbrId, double dist) {
    boolean updated = (dist < distBucket.get(nbrId));
    distBucket.put(nbrId, dist);
    if (updated) { Object[][] resBag = {};
      for (Integer nbr:nbrBucket) resBag.add(nbr,dist+1);
      return resBag;
} } }
WITH SP (srcId, nbrId, dist) AS (
SELECT srcId, -1, 0 FROM graph WHERE srcId = startNode
) UNION ALL UNTIL FIXPOINT by srcId (
SELECT nbr, ArgMin(srcId, distOut).{id, dist}
FROM ( SELECT srcId, SPAgg(nbrId, dist).{nbr, distOut}
       FROM graph, SP WHERE graph.srcId = SP.srcId
       GROUP BY srcId) GROUP BY nbr)
```

### Listing 3: Computing K-means clustering with REX

```
class KMAgg {
  String[] inTypes = {"Integer", "Double", "Double"},
  outTypes={"cid:Integer","xDiff:Double","yDiff:Double"};
  Object[][] update(Object[][] nodeBucket,
  Object[][] centrBucket, int cid, double cx, double cy){
    centrBucket.put(cid, {cx, cy}); Object[][] resBag={};
    for (Integer nid: nodeBucket.keys()) {
      int oldCid = nodeBucket.getCid(nid);
      double oldDist = nodeBucket.getDist(nid);
      double dist = sqrt(sqr(cx-nodeBucket.getx(nid))+
        sqr(cy-nodeBucket.gety(nid)));
      if (oldCid < 0 || dist < oldDist) {
        nodeBucket.putDist(nid, d);
        nodeBucket.putCid(nid, cid);
        resBag.add({cid,nx,ny}, {oldCid,-nx,-ny}); }}
    return resBag;
} }
WITH KM AS (cid, x, y) AS (
SELECT KMSampleAgg(lng,lat).{cid,x,y) FROM geodata
) UNION ALL UNTIL FIXPOINT BY cid (
SELECT cid, avg(xDiff), avg(yDiff)
FROM (SELECT cid, KMAgg(cid, cx, cy).{cid, xDiff, yDiff}
      FROM geodata, KM GROUP BY cid ) GROUP BY cid)
```

## APPENDIX

Listing 2 illustrates a UDA and query for single-source shortest-path. Listing 3 shows two UDAs and a driver query for K-means clustering. *KMAgg* selects the best cluster for each node, adding the node's coordinates to it and subtracting them from the old cluster. For brevity, we omit *InitDist* which assigns initial nodes a distance of 0; *ArgMin*, a general-purpose aggregate returning the identifier with minimum value; and *KMSampleAgg*, which controls how the initial centroids are sampled among the node coordinates.